\documentclass[final]{aipproc}
\usepackage{amssymb}

\layoutstyle{6x9}

\begin{document}

\title{The canonical Gamma-Ray Bursts: long, ``fake''-''disguised'' and ``genuine'' short bursts}

\author{Remo Ruffini}{address={ICRANet, Piazzale della Repubblica 10, 65122 Pescara, Italy.}, altaddress={ICRA and Dip. di Fisica, Univ. di Roma ``La Sapienza'', P.le Aldo Moro 5, 00185 Roma, Italy.}}

\author{Alexey Aksenov}{address={Inst. for Theor. and Experim. Physics, B. Cheremushkinskaya, 25, 117218 Moscow, Russia}}

\author{Maria Grazia Bernardini}{address={ICRANet, Piazzale della Repubblica 10, 65122 Pescara, Italy.}, altaddress={ICRA and Dip. di Fisica, Univ. di Roma ``La Sapienza'', P.le Aldo Moro 5, 00185 Roma, Italy.}}

\author{Carlo Luciano Bianco}{address={ICRANet, Piazzale della Repubblica 10, 65122 Pescara, Italy.}, altaddress={ICRA and Dip. di Fisica, Univ. di Roma ``La Sapienza'', P.le Aldo Moro 5, 00185 Roma, Italy.}}

\author{Letizia Caito}{address={ICRANet, Piazzale della Repubblica 10, 65122 Pescara, Italy.}, altaddress={ICRA and Dip. di Fisica, Univ. di Roma ``La Sapienza'', P.le Aldo Moro 5, 00185 Roma, Italy.}}

\author{Maria Giovanna Dainotti}{address={ICRANet, Piazzale della Repubblica 10, 65122 Pescara, Italy.}, altaddress={ICRA and Dip. di Fisica, Univ. di Roma ``La Sapienza'', P.le Aldo Moro 5, 00185 Roma, Italy.}}

\author{Gustavo De Barros}{address={ICRANet, Piazzale della Repubblica 10, 65122 Pescara, Italy.}, altaddress={ICRA and Dip. di Fisica, Univ. di Roma ``La Sapienza'', P.le Aldo Moro 5, 00185 Roma, Italy.}}

\author{Roberto Guida}{address={ICRANet, Piazzale della Repubblica 10, 65122 Pescara, Italy.}, altaddress={ICRA and Dip. di Fisica, Univ. di Roma ``La Sapienza'', P.le Aldo Moro 5, 00185 Roma, Italy.}}

\author{Gregory Vereshchagin}{address={ICRANet, Piazzale della Repubblica 10, 65122 Pescara, Italy.}}

\author{She-Sheng Xue}{address={ICRANet, Piazzale della Repubblica 10, 65122 Pescara, Italy.}}

\begin{abstract}
The Gamma-Ray Bursts (GRBs) offer the unprecedented opportunity to observe for the first time the blackholic energy extracted by the vacuum polarization during the process of gravitational collapse to a black hole leading to the formation of an electron-positron plasma. The uniqueness of the Kerr-Newman black hole implies that very different processes originating from the gravitational collapse a) of a single star in a binary system induced by the companion, or b) of two neutron stars, or c) of a neutron star and a white dwarf, do lead to the same structure for the observed GRB. The recent progress of the numerical integration of the relativistic Boltzmann equations with collision integrals including 2-body and 3-body interactions between the particles offer a powerful conceptual tool in order to differentiate the traditional ``fireball'' picture, an expanding hot cavity considered by Cavallo and Rees, as opposed to the ``fireshell'' model, composed of an internally cold shell of relativistically expanding electron-positron-baryon plasma. The analysis of the fireshell naturally leads to a canonical GRB composed of a proper-GRB and an extended afterglow. By recalling the three interpretational paradigms for GRBs we show how the fireshell model leads to an understanding of the GRB structure and to an alternative classification of short and long GRBs. 
\end{abstract}

\classification{98.70.Rz}
\keywords      {Gamma-Ray: Bursts}

\maketitle

\section{Brief reminder of the \textit{fireshell} model}\label{model}

\begin{figure}
\centering
\includegraphics[width=0.35\hsize]{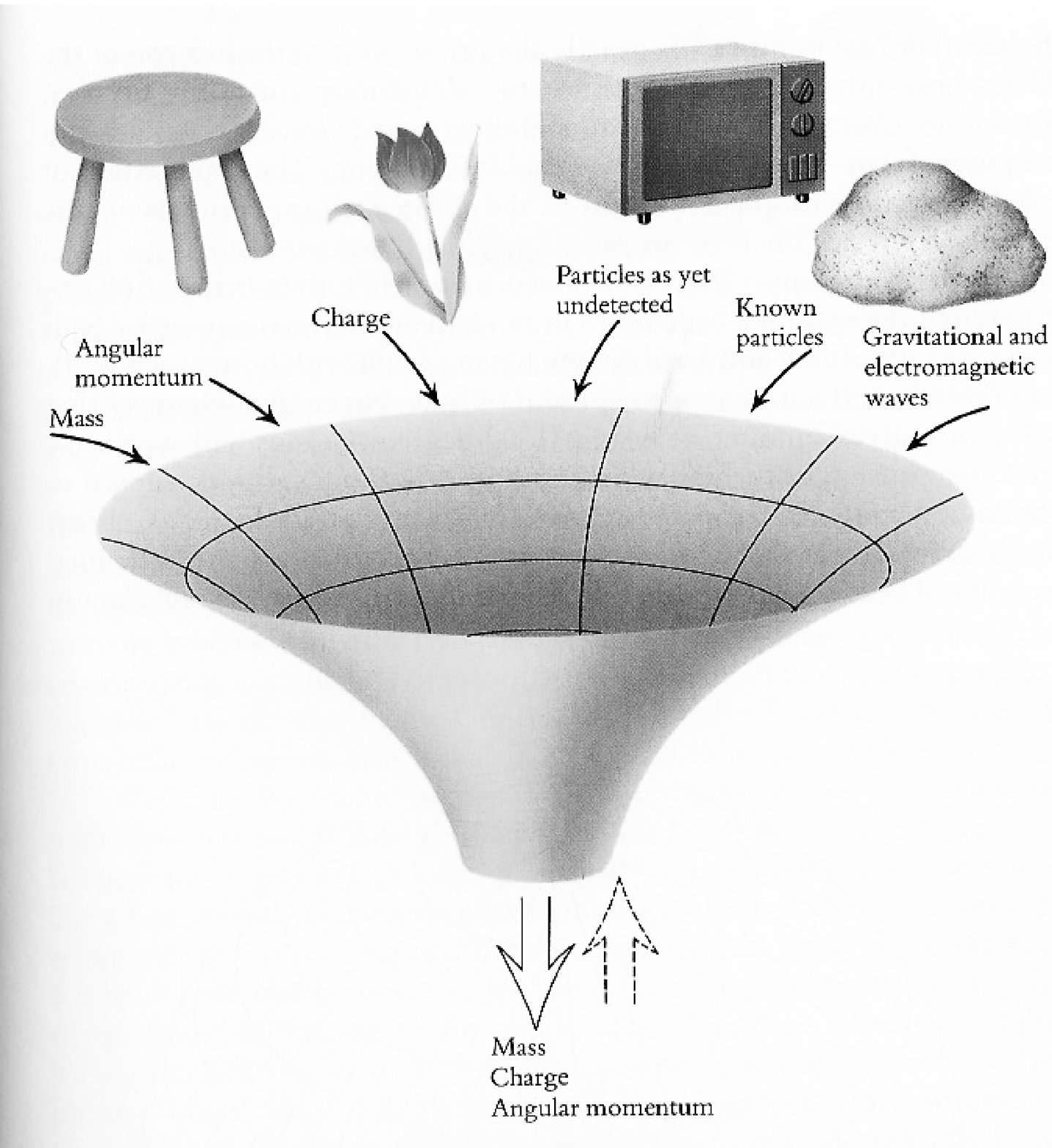}
\includegraphics[width=0.65\hsize]{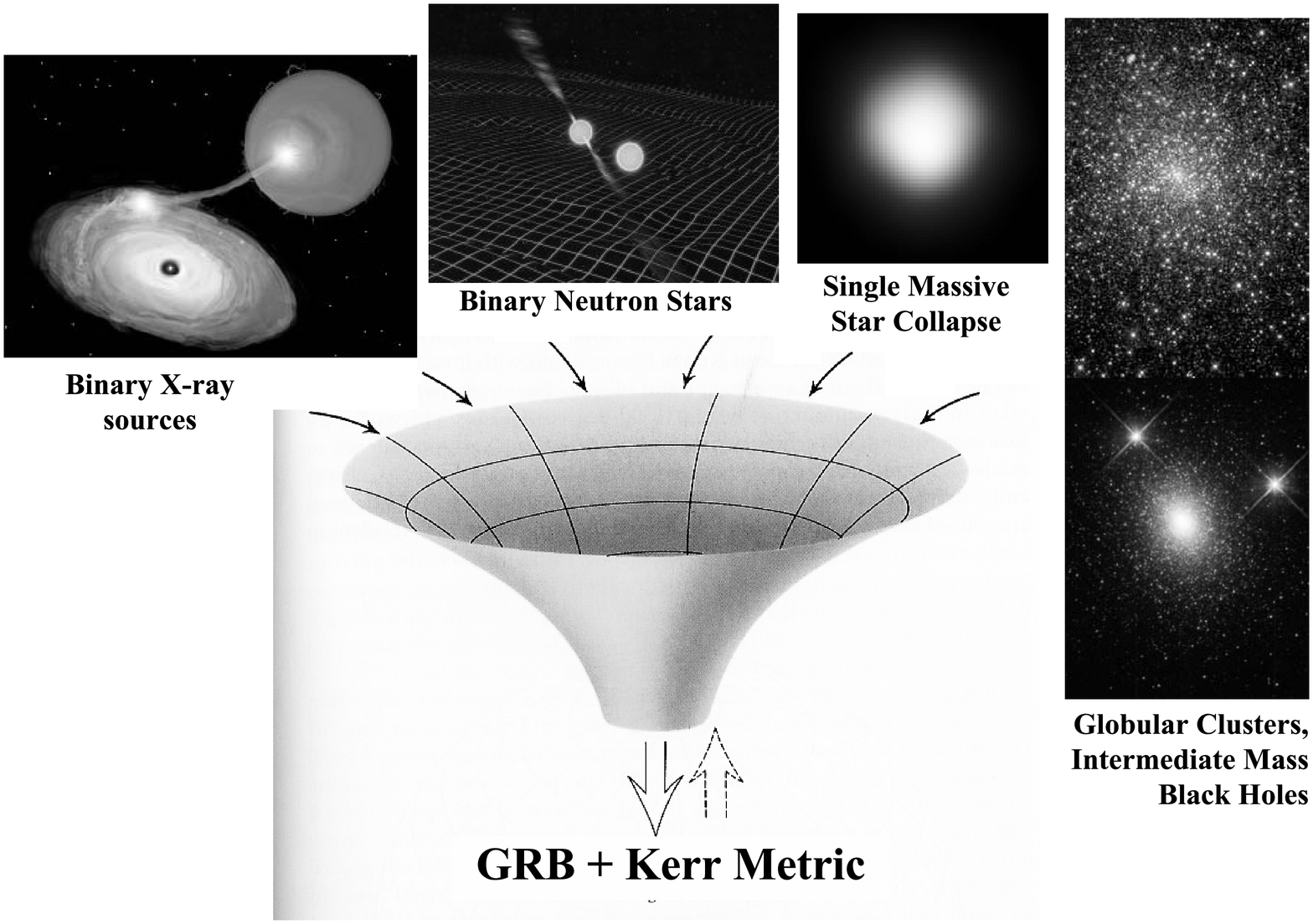}
\caption{\textbf{Left:} The Black Hole Uniqueness theorem \citep[see e.g. Ref.][]{1971PhT....24a..30R}. \textbf{Right:} The GRB uniqueness.}
\label{BH_Uniqueness_1}
\end{figure}

The black hole uniqueness theorem \citep[see Left panel in Fig. \ref{BH_Uniqueness_1} and e.g. Ref.][]{1971PhT....24a..30R} is at the very ground of the fact that it is possible to explain the different GRB features with a single theoretical model, over a range of energies spanning over $6$ orders of magnitude. The fundamental point is that, independently of the fact that the progenitor of the gravitational collapse is represented by merging binaries composed by neutron stars and white dwarfs in all possible combinations, or by a single process of gravitational collapse, or by the process of ``induced'' gravitational collapse, the formed black hole is totally independent from the initial conditions and reaches the standard configuration of a Kerr-Newman black hole (see Right panel in Fig. \ref{BH_Uniqueness_1}). It is well known that pair creation by vacuum polarization process can occur in a Kerr-Newman black hole \citep{1975PhRvL..35..463D,PhysRep}.

We consequently assume, within the fireshell model, that all GRBs originate from an optically thick $e^\pm$ plasma with total energy $E_{tot}^{e^\pm}$ in the range $10^{49}$--$10^{54}$ ergs and a temperature $T$ in the range $1$--$4$ MeV \citep{1998A&A...338L..87P}. Such an $e^\pm$ plasma has been widely adopted in the current literature \citep[see e.g. Refs.][and references therein]{2005RvMP...76.1143P,2006RPPh...69.2259M}. After an early expansion, the $e^\pm$-photon plasma reaches thermal equilibrium with the engulfed baryonic matter $M_B$ described by the dimensionless parameter $B=M_{B}c^{2}/E_{tot}^{e^\pm}$, that must be $B < 10^{-2}$ due to the onset of dynamical instabilities \citep{1999A&A...350..334R,2000A&A...359..855R}. As the optically thick fireshell composed by $e^\pm$-photon-baryon plasma self-accelerate to ultrarelativistic velocities, it finally reaches the transparency condition. A flash of radiation is then emitted. This is the P-GRB \citep{2001ApJ...555L.113R}. Different current theoretical treatments of these early expansion phases of GRBs are compared and contrasted \citep[see e.g. Refs.][and references therein]{brvx06,2008AIPC.1065..219R}. The amount of energy radiated in the P-GRB is only a fraction of the initial energy $E_{tot}^{e^\pm}$. The remaining energy is stored in the kinetic energy of the optically thin baryonic and leptonic matter fireshell that, by inelastic collisions with the CBM, gives rise to a multi-wavelength emission. This is the extended afterglow. It presents three different regimes: a rising part, a peak and a decaying tail. What is usually called ``Prompt emission'' in the current literature mixes the P-GRB with the raising part and the peak of the extended afterglow. Such an unjustified mixing of these components, originating from different physical processes, leads to difficulties in the current models of GRBs, and can as well be responsible for some of the intrinsic scatter observed in the Amati relation \citep{2006MNRAS.372..233A,2008A&A...487L..37G}.

\begin{figure}
\includegraphics[width=\hsize]{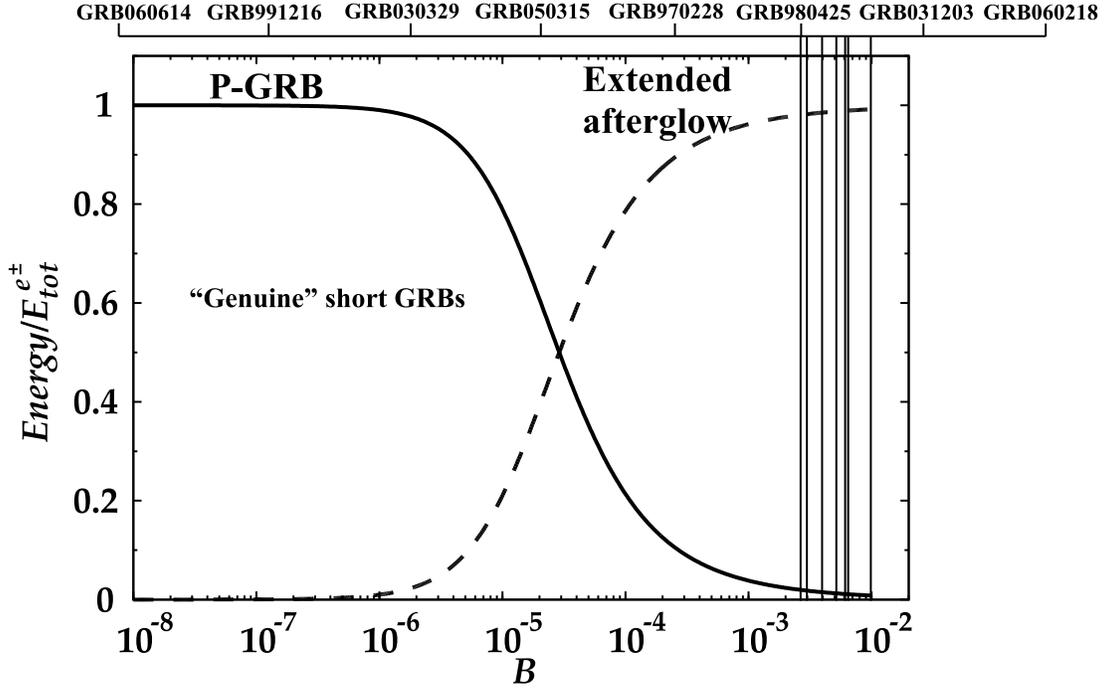}
\caption{Here the energies emitted in the P-GRB (solid line) and in the extended afterglow (dashed line), in units of the total energy of the plasma, are plotted as functions of the $B$ parameter. When $B \lesssim 10^{-5}$, the P-GRB becomes predominant over the extended afterglow, giving rise to a ``genuine'' short GRB. In the figure are also marked the values of the $B$ parameters corresponding to some GRBs we analyzed, all belonging to the class of long GRBs.}
\label{f2}
\end{figure}

At the transparency point, the value of the $B$ parameter rules the ratio between the energetics of the P-GRB and the kinetic energy of the baryonic and leptonic matter giving rise to the extended afterglow. It rules as well the time separation between the corresponding peaks \citep{2001ApJ...555L.113R,2008AIPC.1065..219R}. Within our classification a canonical GRB for baryon loading $B \lesssim 10^{-5}$ has the P-GRB component energetically dominant over the extended afterglow (see Fig. \ref{f2}). In the limit $B \rightarrow 0$ it gives rise to a ``genuine'' short GRB. Otherwise, when $10^{-4} \lesssim B \leq 10^{-2}$, the kinetic energy of the baryonic and leptonic matter, and consequently the emission of the extended afterglow, is dominant with respect to the P-GRB \citep{2001ApJ...555L.113R,2008AIPC.1065..219R,2007A&A...474L..13B,2008AIPC.1065..223B}.

The extended afterglow luminosity in the different energy bands is governed by two quantities associated to the environment. Within the fireshell model, these are the effective CBM density profile, $n_{cbm}$, and the ratio between the effective emitting area $A_{eff}$ and the total area $A_{tot}$ of the expanding baryonic and leptonic shell, ${\cal R}= A_{eff}/A_{tot}$. This last parameter takes into account the CBM filamentary structure \citep{2004IJMPD..13..843R,2005IJMPD..14...97R} and the possible occurrence of a fragmentation in the shell \citep{2007A&A...471L..29D}. In our hypothesis, the emission from the baryonic and leptonic matter shell is spherically symmetric. This allows us to assume, in a first approximation, a modeling of thin spherical shells for the CBM distribution and consequently to consider just its radial dependence \citep{2002ApJ...581L..19R}. The emission process is postulated to be thermal in the co-moving frame of the shell \citep{2004IJMPD..13..843R}. The observed GRB non-thermal spectral shape is due to the convolution of an infinite number of thermal spectra with different temperatures and different Lorentz and Doppler factors. Such a convolution is to be performed over the surfaces of constant arrival time of the photons at the detector \citep[EQuiTemporal Surfaces, EQTSs; see e.g. Ref.][]{2005ApJ...620L..23B} encompassing the whole observation time \citep{2005ApJ...634L..29B}.

\begin{figure}
\includegraphics[width=\hsize,clip]{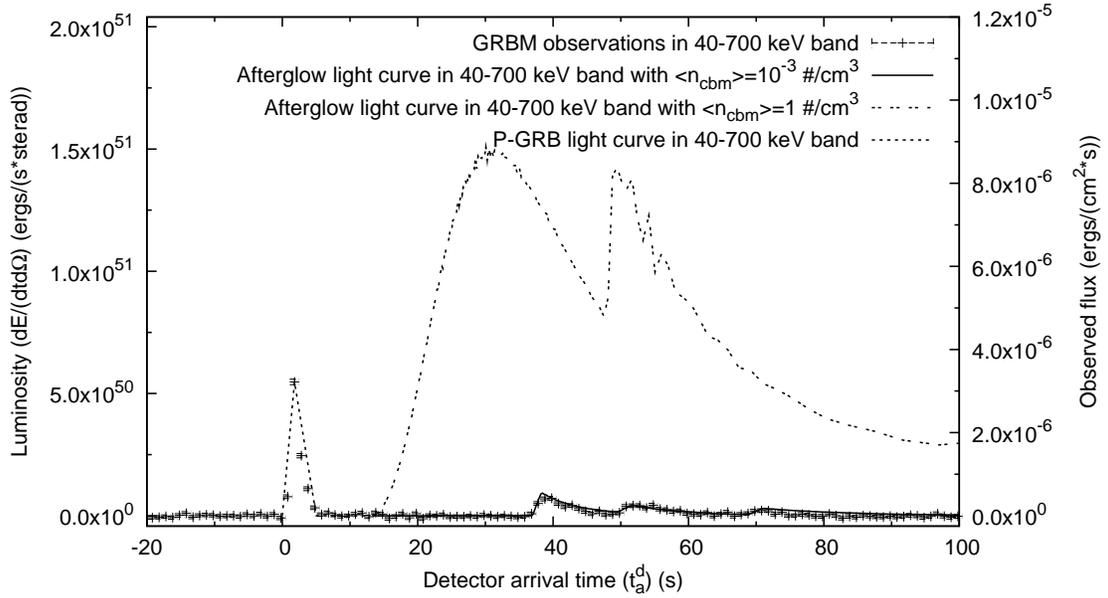}
\caption{The theoretical fit of the \emph{Beppo}SAX GRBM observations of GRB970228 in the $40$--$700$ keV energy band are given. The red line corresponds to an average CBM density $\sim 10^{-3}$ particles/cm$^3$. The black line is the extended afterglow light curve obtained rescaling the CBM density to $\langle n_{cbm} \rangle = 1$ particle/cm$^3$ keeping constant its shape and the values of the fundamental parameters of the theory $E_{e^\pm}^{tot}$ and $B$. The blue line is the P-GRB. Details in \citet{2007A&A...474L..13B}.}
\label{picco_n=1}
\end{figure}

We recently extended the theoretical understanding, within the fireshell model, of a new class of sources, pioneered by \citet{2006ApJ...643..266N}. This class is characterized by an occasional softer extended emission after an initial spikelike emission. The softer extended emission has a peak luminosity smaller than the one of the initial spikelike emission. This has misled the understanding of the correct role of the extended afterglow. As shown in the prototypical case of GRB970228 \citep{2007A&A...474L..13B}, the initial spikelike emission can be identified with the P-GRB and the softer extended emission with the peak of the extended afterglow. Crucial is the fact that the time-integrated extended afterglow luminosity is much larger than the P-GRB one, and this fact unquestionably identifies GRB970228 as a canonical GRB with $B > 10^{-4}$. The consistent application of the fireshell model allowed to compute the CBM porosity, filamentary structure and average density which, in that specific case, resulted to be $n_{cbm} \sim 10^{-3}$ particles/cm$^3$ \citep{2007A&A...474L..13B}. This explained the peculiarity of the low extended afterglow peak luminosity and of its much longer time evolution. These features are not intrinsic to the progenitor nor to the black hole, but they uniquely depend on the peculiarly low value of the CBM density, typical of galactic halos. If the same total energy, baryon loading and CBM distribution as in GRB970228 is taken and the CBM density profile is rescaled by a constant numerical factor in order to raise its average value from $10^{-3}$ to $1$ particles/cm$^3$, it is obtained a GRB with a much larger extended afterglow peak luminosity and a much reduced time scale. Such a GRB would appear a perfect traditional ``long'' GRB following the current literature \citep[see Fig. \ref{picco_n=1} and Ref.][]{2007A&A...474L..13B}. This has led us to expand the traditional classification of GRBs to three classes: ``genuine'' short GRBs, ``fake'' or ``disguised'' short GRBs, and all the remaining ``canonical'' ones.

\section{The ``fireshell'' model and GRB progenitors}\label{progenitors}

``Long'' GRBs are traditionally related in the current literature to the idea of a single progenitor, identified as a ``collapsar'' \citep{1993ApJ...405..273W}. Similarly, short GRBs are assumed to originate from binary mergers formed by white dwarfs, neutron stars, and black holes in all possible combinations \citep[see e.g. Refs.][and references therein]{2005RvMP...76.1143P,2006RPPh...69.2259M}. It has been also suggested that short and long GRBs originate from different galaxy types. In particular, short GRBs are proposed to be associated with galaxies with low specific star forming rate \citep[see e.g. Ref.][]{2009ApJ...690..231B}. Some evidences against such a scenario have been however advanced, due to the small sample size and the different estimates of the star forming rates \citep[see e.g. Ref.][]{2008arXiv0803.2718S}. However, the understanding of GRB structure and of its relation to the CBM distribution, within the fireshell model, leads to a more complex and interesting perspective than the one in the current literature.

\begin{figure}
\includegraphics[width=\hsize,clip]{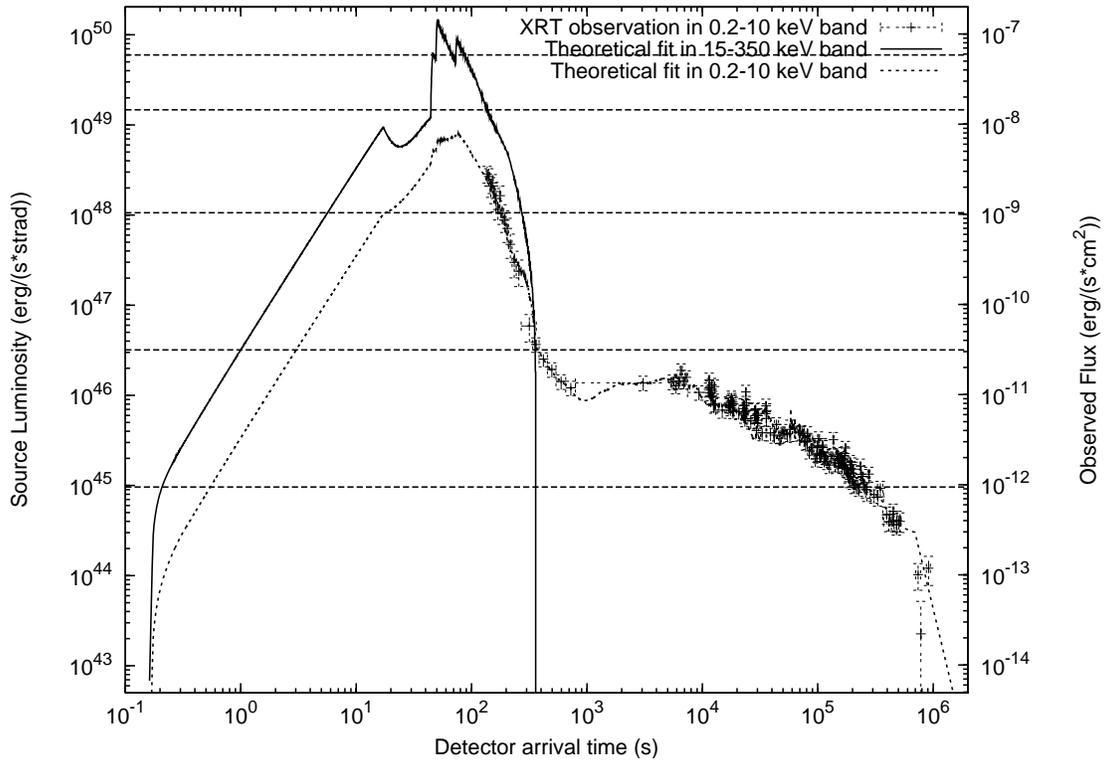}
\caption{The theoretical light curves in the $15-150$ keV (solid line) and $0.2-10$ kev (dotted line) energy bands compared with XRT observations of GRB 050315 \citep{2006ApJ...638..920V}. The horizontal dashed lines correspond to different possible instrumental thresholds. It is clear that long GRB durations are just functions of the observational threshold. Details in \citet{Venezia_Orale}.}
\label{global_th}
\end{figure}

The first general conclusion of the ``fireshell'' model \citep{2001ApJ...555L.113R} is that, while the time scale of ``short'' GRBs is indeed intrinsic to the source, this does not happen for the ``long'' GRBs: their time scale is clearly only a function of the instrumental noise threshold. This has been dramatically confirmed by the observations of the Swift satellite \citep[see Fig. \ref{global_th} and Ref.][]{Venezia_Orale}. Among the traditional classification of ``long'' GRBs we distinguish two different sub-classes of events, none of which originates from collapsars.

The first sub-class contains ``long'' GRBs particularly weak ($E_{iso} \sim 10^{50}$ erg) and associated with SN Ib/c. In fact, it has been often proposed that such GRBs, only observed at smaller redshift $0.0085 < z < 0.168$, form a different class, less luminous and possibly much more numerous than the high luminosity GRBs at higher redshift \citep{2006Natur.442.1011P,2004Natur.430..648S,2007ApJ...658L...5M,2006AIPC..836..367D}. Therefore in the current literature they have been proposed to originate from a separate class of progenitors \citep{2007ApJ...662.1111L,2006ApJ...645L.113C}. Within our ``fireshell'' model, they originate as well in a separate class of progenitors composed by binary systems formed by a neutron star, close to its critical mass, and a companion star, evolved out of the main sequence. They produce GRBs associated with SNe Ib/c, via the ``induced gravitational collapse'' process \citep{2001ApJ...555L.117R}. The particularly low luminosity observed in these sources is then explained by the formation of a black hole with the smallest possible mass: the one formed by the collapse of a just overcritical neutron star \citep{Mosca_Orale,2007A&A...471L..29D}.

A second sub-class of ``long'' GRBs originates from merging binary systems, formed either by two neutron stars or a neutron star and a white dwarf. A prototypical example of such systems is GRB970228. The binary nature of the source is inferred by its migration from its birth location in a star forming region to a low density region within the galactic halo, where the final merging occurs \citep{2007A&A...474L..13B}. The location of such a merging event in the galactic halo is indeed confirmed by optical observations of the GRB970228 afterglow  \citep{1997Natur.387R.476S,1997Natur.386..686V}. The crucial point is that, as recalled above, GRB970228 is a ``canonical'' GRB with $B > 10^{-4}$ ``disguised'' as a short GRB.

If the binary merging would occur in a region close to its birth place, with an average density of $1$ particle/cm$^3$, the GRB would appear as a traditional high-luminosity ``long'' GRB, of the kind currently observed at higher redshifts (see Fig. \ref{picco_n=1}), similar to, e.g., GRB050315 \citep{2006ApJ...645L.109R}.

Within our approach, therefore, there is the distinct possibility that all GRB progenitors are formed by binary systems, composed by neutron stars, white dwarfs, or stars evolved out of the main sequence, in different combinations.

The case of the ``genuine'' short GRBs is currently being examined within the ``fireshell'' model.

\section{Open issues}\label{open_issues}

The ``fireshell'' model addresses mainly the $\gamma$ and X-ray emission, which are energetically the most relevant part of the GRB phenomenon. The model allows a detailed identification of the fundamental three parameters of the GRB source: the total energy, the baryon loading, as well as the CBM density, filamentary structure and porosity. The knowledge of these phenomena characterizes the region surrounding the black hole up to a distance which in this source reaches $\sim 10^{17}$--$10^{18}$ cm. When applied, however, to larger distances, which corresponds to the latest phases of the X-ray afterglow, since the ``plateau'' phase, the model reveals a different regime which has not yet been fully interpreted in its astrophysical implications. To fit the light curve in the soft X-ray regime for $r > 4 \times 10^{17}$ cm, we must appeal to an anomalous enhancement of about six orders of magnitude in the ${\cal R}$ factor. This corresponds to a much more diffuse CBM structure, with a smaller porosity, interacting with the fireshell. This points to a different leading physical process during the latest X-ray afterglow phases. When we turn to the optical, IR and radio emission, the fireshell model leads to a much smaller flux than the observed one, especially for $r \sim 10^{17}$--$10^{18}$ cm. It is well conceivable that the synchrotron radiation \citep{1997ApJ...476..232M} in these latest phases becomes equally relevant, if not predominant, with respect to the thermal emission which we consider and which dominates the prompt emission. Indeed, though the optical, IR, and radio luminosities have a minority energetic role, they may lead to the identification of crucial parameters and new phenomena occurring in the interaction of the fireshell with the CBM at a distance on the order of a light year away from the newly born black hole. These processes must be considered with the maximum attention.

\section{Conclusions}

These results lead to three major new possibilities:
\begin{itemize}
\item The majority of GRBs declared as shorts \citep[see e.g. Ref.][]{2005Natur.437..822P} are likely ``disguised'' short GRBs, in which the extended afterglow is below the instrumental threshold.
\item Recently, there has been a crucial theoretical physics result, showing that the characteristic time constant for the thermalization for an $e^\pm$ plasma is on the order of $10^{-13}$ s \citep{2007PhRvL..99l5003A}. Such a time scale still applies for an $e^\pm$ plasma with a baryon loading on the order of the one observed in GRBs \citep{PRD}. The shortness of such a time scale, as well as the knowledge of the dynamical equations of the optically thick phase preceding the P-GRB emission \citep{brvx06}, implies that the structure of the P-GRB is a faithful representation of the gravitational collapse process leading to the formation of the black hole \citep{2005IJMPD..14..131R}. In this respect, it is indeed crucial that the Swift data on the P-GRB observed in GRB060614 \citep{2006Natur.444.1044G,2007A&A...470..105M,2008arXiv0810.4855C} appear to be highly structured all the way to time scale of $0.1$ s. This opens a new field of research: the study of the P-GRB structure in relation to the nature of the progenitors leading to the process of gravitational collapse generating the GRB.
\item As recalled above, if indeed the binary nature of the progenitor system and the peculiarly low CBM density $n_{cbm} \sim 10^{-3}$ particles/cm$^3$ will be confirmed for all ``fake'' or ``disguised'' GRBs, then it is very likely that the traditionally ``long'' high luminosity GRBs at higher redshift also originates from the merging of binary systems formed by neutron stars and/or white dwarfs occurring close to their birth location in star forming regions with $n_{cbm} \sim 1$ particle/cm$^3$ (see Fig. \ref{picco_n=1}).
\end{itemize}

\end{document}